\documentclass[aps,pra,twocolumn,floats,epsfig,showpacs]{revtex4-1}%
\usepackage{bbm}
\usepackage{amssymb}
\usepackage{ifpdf}
\usepackage{color}
\usepackage{graphicx}
\usepackage{amsmath}
\usepackage{amsfonts}
\usepackage{hyperref}
\usepackage{dcolumn}
\usepackage{bm}
\usepackage{xcolor}
\usepackage{subfigure}
\usepackage{verbatim}%
\begin{document}
\title{Probing the flat band of optically trapped spin-orbital-coupled Bose gases using Bragg spectroscopy }
\author{Wu Li$^{1}$, Lei Chen$^{1}$, Zhu Chen$^{2}$, Ying Hu$^{3}$, Zhidong Zhang$^{1}$, and Zhaoxin
Liang$^{1}$}\email{E-mail: zhxliang@imr.ac.cn}
\affiliation{$^{1}$ Shenyang National Laboratory for Materials Science, Institute of Metal
Research, CAS, Wenhua Road, 72, Shenyang, China}
\affiliation{$^{2}$ National Key Laboratory of Science and Technology on Computational
Physics, Institute of Applied Physics and Computational Mathematics, Beijing
100088, China}
\affiliation{$^{3}$ Institute for Quantum Optics and Quantum Information of the Austrian Academy of Sciences, A-6020 Innsbruck, Austria}

\begin{abstract}
Motivated by the recent efforts in creating flat bands in ultracold atomic
systems, we investigate how to probe a flat band in an
optically trapped spin-orbital-coupled Bose-Einstein condensate using Bragg spectroscopy. We find that the excitation spectrum and the dynamic structure factor of the condensate are dramatically altered when the band structure exhibits various levels of flatness. In particular, when the band exhibits perfect flatness around the band minima corresponding to a near-infinite effective mass, a quadratic dispersion emerges in the low-energy excitation spectrum; in sharp contrast, for the opposite case when an ordinary band is present,  the familiar linear dispersion arises. Such linear-to-quadratic crossover in the low-energy spectrum presents a striking manifestation of the transition of an ordinary band into a flat band, thereby allowing a direct probe of the flat band by using Bragg spectroscopy.

\end{abstract}

\maketitle

\section{Introduction}
There have been intensive efforts in realizing flat bands in various context of condensed-matter \cite{Fbcmp} and atomic physics \cite{Yao2012,You2012,Goldman2014,Choudhury2014, Lin2014}. The motivation behind this search is twofold. First, a flat band, whose kinetic energy is highly quenched compared to the scale of interactions, possesses macroscopic level degeneracy, and as a result, interactions play a dominant role in affecting the system that has given rise to many interesting quantum phases \cite{You2012}. Second, even more challenging is to create topological flat bands with nonzero Chern number \cite{Bergholtz2013}, which can open a new avenue for engineering a fractional
topological quantum insulator \cite{Fbcmp,Yao2012} without Landau levels prompted by the analogy to
Landau levels \cite{BookLandau} in condensed matter physics. Motivated by the ongoing interests in creating flat bands in ultracold atomic systems\cite{Yao2012,You2012,Goldman2014,Choudhury2014,Lin2014,Bergholtz2013,Zhang2013}, we address below the problem of how to probe an arising isolated flat band in an optically trapped spin-orbital-coupled (SOC)  Bose-Einstein condensate (BEC) \cite{Zhang2013} by using Bragg spectroscopy.

The key ingredient of our work consists in investigating how the
excitation spectrum and dynamic structure factor of the system change when the band structure varies its flatness. Our main results are as follows: (1) a quadratic dispersion $\epsilon(k)\sim k^2$ emerges in the low-energy excitation spectrum, if the band is perfectly flat in the vicinity of energy band minima (corresponding to an infinite effective mass); contrasting sharply, in the opposite case when the BEC has an ordinary band, the familiar linear dispersion relation $\epsilon(k)\sim k$ is found; (2) the static structure factor $S(k)$ exhibits a crossover from linear ($S(k)\sim k$) to quadratic relation ($S(k)\sim k^2$) in the momenta, when the band transforms from the ordinary into the flat band. Moreover, by relating the flatness of the band with the effective mass at the band minima and by using Feynman's relation $\epsilon(k)=\epsilon_0(k)/S(k)$ \cite{PineBook,He2012,Explain}, we are able to directly connect the emerging quadratic dispersion in a perfect flat band case with the vanishingly small kinetic energy in the single-particle energy $\epsilon_0(k)$.

\begin{figure}[ptb]
\centering
%Requires \usepackage{graphicx}
\includegraphics[width=8.0cm]{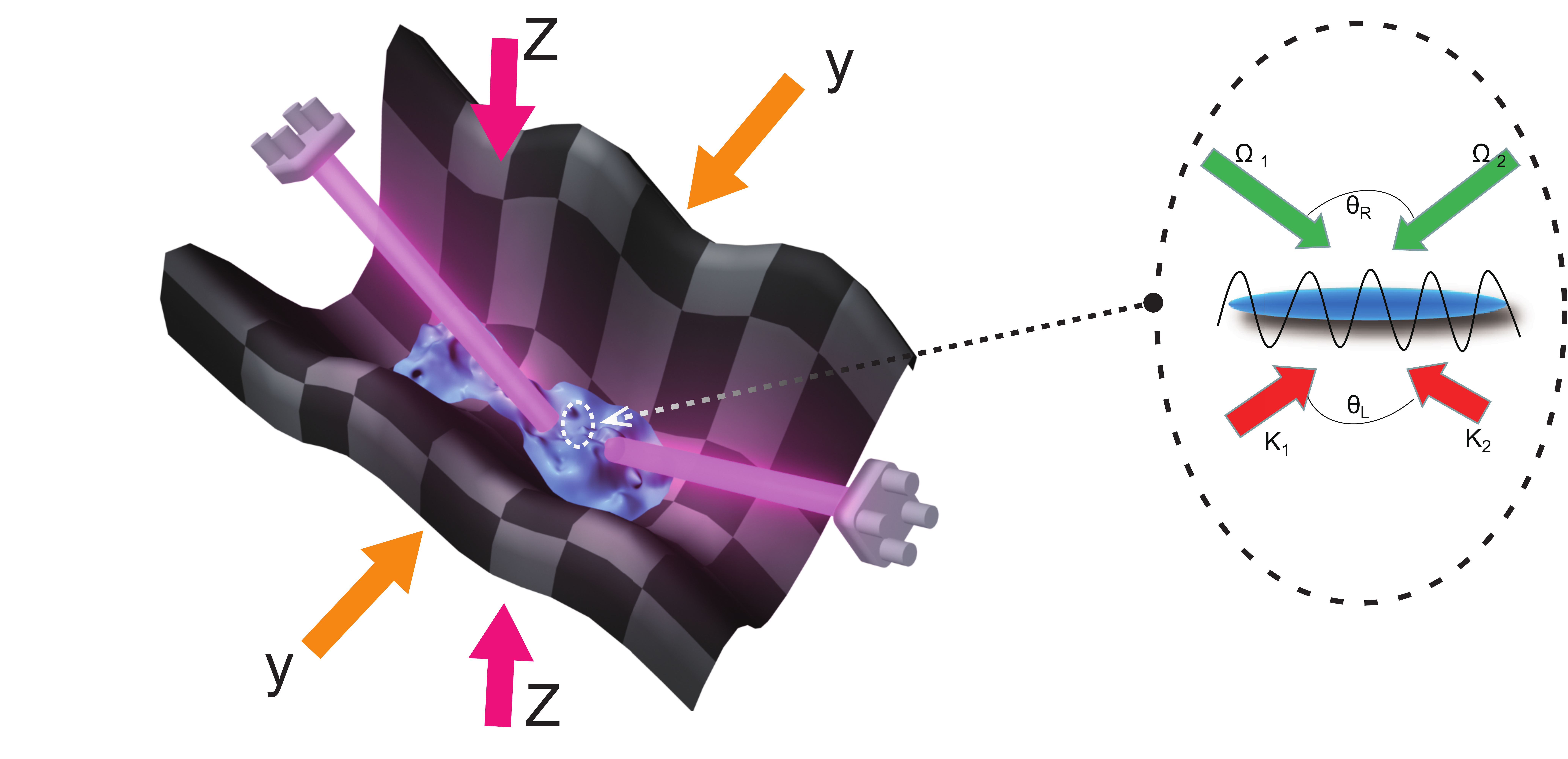}
\caption{(color online).On the left Bragg spectroscopy probes a single-particle flat band generated by an optically trapped spin-orbit-coupled Bose gas.
Right panel: schematic setup for implementing a flat band realized in 1D spin-orbit-coupled Bose gas proposed by Ref. \cite{Zhang2013}. Here, $\Omega_1$ and $\Omega_2$ are the Rabi frequencies of the Raman lasers for generating SO coupling. The interference of other two counterpropagating laser beams labeled by ${\bf k}_1$ and ${\bf k}_2$ generates an optical lattice. Bragg spectroscopy can be described by the dynamic structure factor of the model system.}%
\label{sketch}%
\end{figure}

The setting we consider to probe a flat band in a quasi-one-dimensional BEC with spin-orbit coupling(SOC) trapped in an optical lattice along $x$-direction is illustrated in Fig. \ref{sketch}. In addition, we strongly confine the BEC in both $y$- and $z$- directions such that the dynamics of the model system is effectively restricted to one dimension. Experimentally, the setting in Fig. \ref{sketch} can be realized by combining 
Bragg spectroscopy \cite{BraggOL1,BraggOL2} and SOC \cite{SOCRev1,SOCRev2,SOCRev3} that are available in both
BECs \cite{SOCBose} and Fermi gases \cite{SOCFermi}. In particular, the one-dimensional(1D) SOC with equal Rashba and Dresselhaus contributions considered in this work has been implemented in Ref. \cite{SOCBose} by coupling two internal states of atoms $^{87}Rb$ via Raman lasers. Very recently, Bragg spectroscopy has been employed to reveal the structure of the excitation spectrum in a BEC with SOC in the free space \cite{Ji,Ha,Khamehchi}. In particular, the measurement of the static structure factor combined with  Feynman's relation \cite{Ji,Ha,Khamehchi} has allowed the experimental verification of the emerging roton-maxon dispersion in these systems. Building on this experimental progress in applying  Bragg spectroscopy in a free BEC with SOC, we propose that the Bragg spectroscopy in an  optically trapped BEC with SOC in quasi-one-dimension can help reveal the linear-to-quadratic transition in the low-energy spectrum predicted in this work and therefore provide a direct experimental probe of a flat band. Theoretically, the Gross-Pitaevskii equation (GPE) has been shown to describe well, at the mean-field level, both the static and the dynamic properties of a BEC with SOC \cite{Zhang2013,GPE1,GPE2,GPE3,GPE4}. The validity of
the GPE can be tested a \emph{posteriori} by evaluating the quantum depletion of the condensate. For a more rigorous proof of validity of GPE, we refer to the Supplemental Material in Ref. \cite{Ramos}.

This paper is organized as follows. In Sec. \ref{FB},  we shall begin with briefly describing the model system in which a flat band can arise following Ref.~\cite{Zhang2013}. Then, in Sec. \ref{BS} we show how Bragg spectroscopy can present as an efficient tool to quantitatively probe the presence of a flat band. Finally, Sec. \ref{CO} is devoted to the discussion of observing the described phenomena in a possible experimental parameter regime and a summary of our work.

\begin{figure*}
	\includegraphics[width=1.0\textwidth]{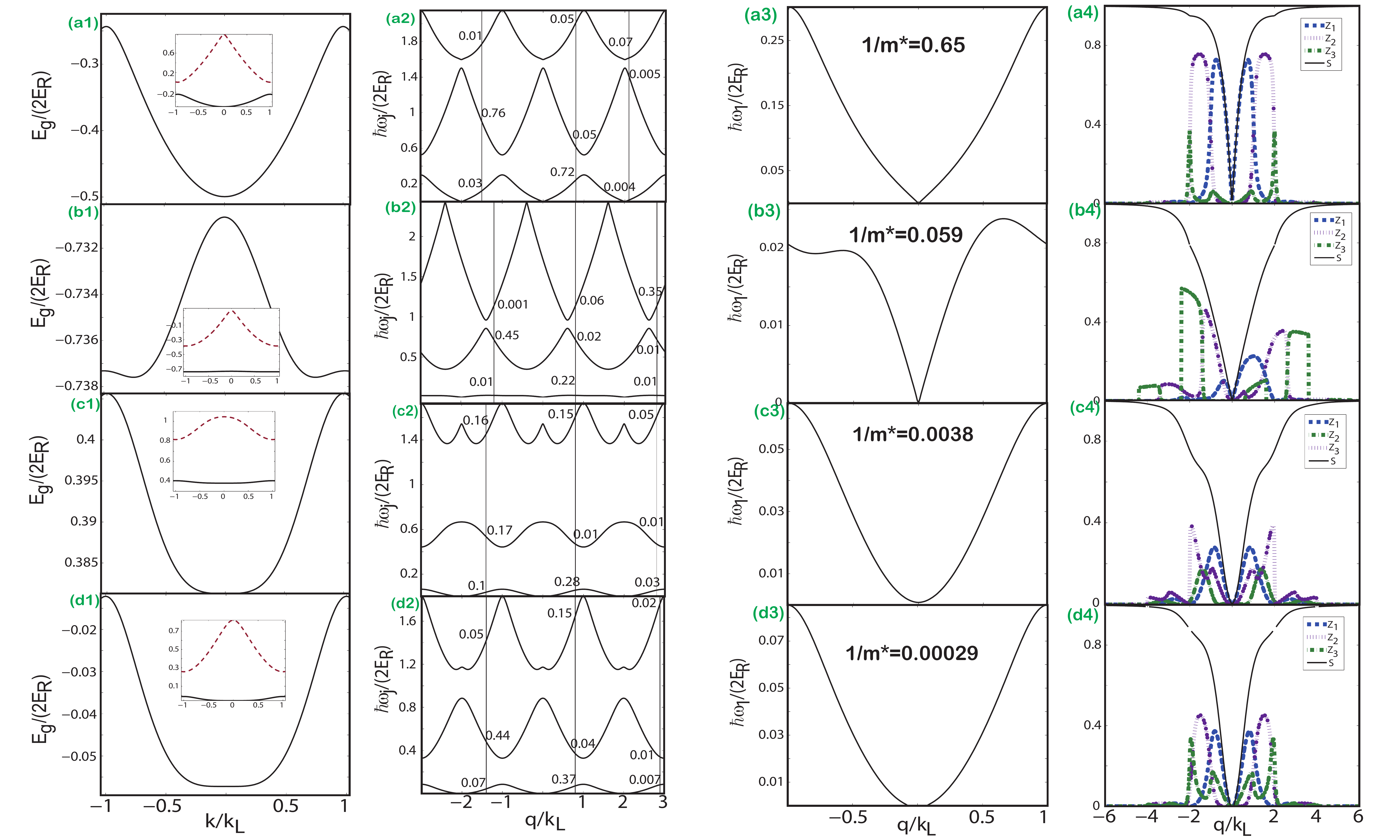}
\caption{(color online) ($a1$-$d1$) are the lowest Bloch band of an optically-trapped SO coupled BEC (in the insert, the lowest two bloch bands are plotted); ($a2$-$d2$) are the lowest three Bogoliubov bands $\omega_{j}$ and the vertical lines represent the excitation strengths $Z_j$ ($j=1,2,3$) toward the first three bands for $q=0.8k_{L}$, $q=-1.2k_{L}$, $q=2.8k_{L}$ respectively. ($a3$-$d3$) are the lowest excitation spectrum while ($a4$-$d4$) are the excitation strength ($Z_{1,2,3}$) and static structure factor ($S$) via a given transferred momentum $q$. Here, $m^*$ labels the effective mass at the energy
minima $k_{min}$. The parameters are given as follows: $c$=0.0500  and ($a1,a2,a3,a4$) $\gamma=0.5000$, $V_{0}=0.6000$,and $\Omega$=0.8000; ($b1,b2,b3,b4$) $\gamma=1.0500$, $V_{0}=1.0000$, and $\Omega=1.1500$;  ($c1,c2,c3,c4$) $\gamma=0.6100$, $V_{0}=2.0000$, and $\Omega=0.3700$; ($d1,d2,d3,d4$) $\gamma=0.7000$, $V_{0}=1.0000$, and $\Omega=0.5005$.}
\label{band}
\end{figure*}

\section{Emerging flat bands in a spin-orbit-coupled BEC}  \label{FB}
The system considered in this work is illustrated in Fig. \ref{sketch}, which consists of a BEC with 1D SOC that is trapped in a strongly anisotropic lattice potential. The transverse lattice confinement in the $y$- and $z$-directions is sufficiently strong to freeze the atomic motion in these directions, allowing the atomic tunneling only in the $x$-direction \cite{Ying2011}. This realizes an optically trapped quasi-1D BEC with 1D SOC in the $x$-direction, which can be well described by the GPE \cite{Zhang2013,GPE1,GPE2,GPE3,GPE4},
\begin{equation}\label{Ham}
i\hbar\frac{\partial\Psi}{\partial t}=\left( {H}_{0}+{H}_{\textrm{int}}\right) \Psi,
\end{equation}
with $\Psi=\left( \psi_{\uparrow},\psi_{\downarrow}\right) ^{T}$ being the
two-component condensate wave functions. The Hamiltonian $H_0$ describes non-interacting bosons in a 1D optical lattice with SOC, reading
\begin{equation}
{H}_{0}=\frac{p_{x}^{2}}{2m}+\gamma p_{x}\sigma_{z}+\Omega\sigma_{x}+V_{0}\times E_R\sin^{2}\left(  k_{L}x\right),\label{SingleH}
\end{equation}
where $m$ is the bare atom mass, $\sigma_{x}$ and $\sigma_{z}$ are the $x$- and $z$-component of Pauli matrices, $\Omega$ is the Rabi frequency for generating SOC, $\gamma=2\pi\hbar\sin\left(  \theta_{R}/2\right)/(\lambda_{R} m)$ with $\lambda_R$ being the wavelength of the two Raman lasers and $\theta_{R}$ the angle between the lasers; and $V_0$ labels the lattice strength in the unit of the recoil energy
$E_{R}=\hbar^2 k_{L}^{2}/2m$, with $k_{L}$ the wave vector of the lasers creating the optical lattice. The Hamiltonian $H_{\textrm{int}}$ describes the hard-core interaction between bosonic atoms, which can be generally written as
\begin{equation}
{H}_{\textrm{int}}=\int dx\left( g_{11}n_{\uparrow}^{2}+g_{22}n_{\downarrow}^{2}
+2g_{12}n_{\uparrow}n_{\downarrow}\right)，,
\end{equation}
where $n_{\uparrow}=|\psi_{\uparrow}|^2$ and $n_{\downarrow}=|\psi_{\downarrow}|^2$ are the two-component condensate densities, and $g_{ij}=4\pi\hbar^2 a_{ij}/m$ ($i,j$=$1$ or $2$) is the coupling constant, with $a_{ij}$ the $s$-wave scattering length.
In this work, we limit ourselves to the case when $g_{11}=g_{22}=g_{12}=g=4\pi\hbar^2a/m>0$; in this regime, the striped phase will not appear in the ground state. For later convenience,  we rescale GP Eq. (\ref{Ham})  into the dimensionless form by introducing $x\rightarrow k_L x$, $t\rightarrow (2E_R/\hbar) t$, $\gamma\rightarrow \gamma/(\hbar k_L/m)$, $\Omega\rightarrow\Omega/2E_R$, and
the dimensionless interaction coefficient $c = \sqrt{\omega_y \omega_z}k_L a N/E_R$ with $N$ the atom number in one unit cell, and $\omega_y$ and $\omega_z$ the trapping frequencies in the transverse directions.

The physics of an optically trapped quasi-1D BEC with SOC is governed by the interplay among four parameters: the SOC parameters $\gamma$ and $\Omega$, lattice strength $V_0$ and interaction $c$. Crucial to the emergence of flat band in such systems, as pointed out in Ref. \cite{Zhang2013}, is the interplay between the SOC parameters ($\gamma$ and $\Omega$) and the lattice strength ($V_0$). The basic mechanism can be intuitively described using the single-particle picture \cite{Zhang2013}: (i) Without the interaction ($c=0$) and the optical
potential ($V_0=0$), the single-particle Hamiltonian $H_0$ can be cast into a dimensionless form \begin{equation}
{H}_0=\left(
\begin{array}
[c]{cc}%
\frac{k^{2}}{2}+\gamma k & \Omega\\
\Omega & \frac{k^{2}}{2}-\gamma k
\end{array}
\right) , \label{H00}
\end{equation}
which has two energy bands $\mu_{\pm}\left(  k\right)  =k^{2}/2
\pm\sqrt{\gamma^{2}k^{2}+\Omega^{2}}$ separated by a band gap $2\Omega$ at $k=0$. (ii) When an optical lattice ($V_0\neq 0$) is added to Hamiltonian (\ref{H00}), a second band gap will open at the edge of Brillouin Zone, with the magnitude of the gap being dependent on $V_0$. (iii) By engineering (via tuning $\gamma$, $\Omega$ and $V_0$) the magnitude of both gap, a flat band can be realized.  Strikingly,  the existence of flat bands stays robust against the mean-field interaction in the BEC according to Ref. \cite{Zhang2013}.

In Figs. \ref{band} (a1)-(d1), we have plotted the lowest Bloch bands $E_g(k)$ for various choices of the SOC parameters ($\gamma$ and $\Omega$) and lattice strength $V_0$ by numerically solving
Eq. (\ref{Ham})  with fixed interaction parameter $c$  (detailed numerical method can be found in Ref. \cite{Liang,LeiChenJLTP,LeiChenEPJD}). The presence of flat band is manifest to the eye (see Fig.\ref{band} (b1), (c1) and (d1)), as compared to an ordinary band (see Fig.\ref{band} (a1)). Quantitatively, the global flatness of the bands can be measured by the ratio $W$ between the bad gap and the band width \cite{Zhang2013}.

When the model BEC system is probed by the Bragg spectroscopy,  it is the excitation near the band minima $k_{min}$ that is addressed in the linear perturbation regime. Therefore, we expect the {\it local flatness} at $k_{min}$ to be directly probed in Bragg spectroscopy, rather than the global flatness measured by $W$.

In order to characterize the local flatness near the band minima, we have calculated the effective mass $m^*(k_{min})$ for various bands (in this work, whenever we use the notation $m^*$, we refer to the effective mass evaluated at $k_{min}$). Our calculation shows that an ordinary band has $m^*\sim 1$ (e.g. $1/m^*=0.65$ in Fig. \ref{band} (a3)), while in comparison, the flat band has much larger effective mass $m^*>>1$ as expected (see Figs. \ref{band} (b3)-(d3)).  Interestingly, $m^*$ also varies sharply for various flat band, such that we can further discriminate between the sectional flat band (see Figs. \ref{band} (c1)-(d1)) and the global flat band (see Figs.\ref{band} (b1)), the former having much bigger effective mass $m^*$ than the latter. In other words, the sectional band is locally much flatter near $k_{min}$ than the global flat band, even though its global flatness measured by $W$ can be actually smaller. Figure \ref{band} (b1) presents a typical globally flat band, which has $1/m^*=0.059$; whereas, Figs.\ref{band} (c1)  and (d1) present two sectional flat band, which have $1/m^*=0.0038$ and $1/m^*=0.00029$, respectively. Noticing that $m^*\rightarrow \infty$ for the sectional band in Fig. \ref{band} (d1), we shall call it as a perfect flat band. As we shall see, the excitation behaviour of the model BEC can alter significantly when $m^{*}$ and the flatness of band changes.

\section{Probing flat bands using Bragg spectroscopy} \label{BS}

We now discuss how the flatness of a band in a SOC BEC (see Fig. \ref{sketch}) can be revealed in Bragg spectroscopy.Bragg spectroscopy consists here in generating a density perturbation to the model system by using two Bragg laser beams that have momenta ${\bf k}_{1,2}$ and a frequency difference $\omega$ ($\omega$ is much smaller than their detuning from an atomic resonance
\cite{BraggOL1, BraggOL2}). The linear perturbation is described by the Hamiltonian $V_{1}=\frac{V}{2}\left[ \rho^{\dag}_{\mathbf{q}%
}e^{-i\omega t}+\rho_{\mathbf{-q}}e^{+i\omega t}\right] $, where $\rho
_{\mathbf{q}}=\sum_{j}e^{i\mathbf{q}\cdot\mathbf{r}_{j}/\hbar}$ is the
Fourier transformed one-body density operator, and $\mathbf{q}=\mathbf{k}_{1}-\mathbf{k}_{2}$ is the probe momenta. Right after the perturbation, the dynamical structure factor \cite{Kurn,Zambelli} is probed, which is written as
\begin{equation}
S\left( \mathbf{q},\omega\right) = \sum_{e}|\langle e|\rho^{\dag}_{\mathbf{q}}|0\rangle|^{2}
\delta\left( \omega-(E_{e}-E_{g})/\hbar \right),
\end{equation}
 with $|0\rangle$ ($|e\rangle$) being the ground (excited) state having the energy $E_{g}$ ($E_{e}$).  From the dynamic structure factor, the excitation spectrum can then be extracted \cite{LeiChenJLTP,LeiChenEPJD,Zambelli}.

Let us calculate the excitation spectrum and the dynamic structure factor $S(q,\omega)$ of the model system for various band structures, from an ordinary band to a perfect flat band.  For this purpose, we apply the Bogoliubov
theory \cite{LeiChenJLTP,Menotti} to Eq. (\ref{Ham}) and decompose the condensate wave function
$(\psi_{\uparrow}, \psi_{\downarrow})^{T}$ into the ground state wave function
$(\phi_{\uparrow 0}, \phi_{\downarrow 0})^{T}$ and a small fluctuating term reading
\begin{equation}\label{Psigp}
\begin{pmatrix}
\psi_{\uparrow}\\
\psi_{\downarrow}%
\end{pmatrix}
= e^{-i\mu t}\left[
\begin{pmatrix}
\phi_{\uparrow 0}\\
\phi_{\downarrow 0}%
\end{pmatrix}
+
\begin{pmatrix}
u_{\uparrow}(x)\\
u_{\downarrow}(x)
\end{pmatrix}
e^{-i\omega t} +\begin{pmatrix}
v^*_{\uparrow}(x)\\
v^*_{\downarrow}(x)
\end{pmatrix}
e^{i\omega t}\right].
\end{equation}
By substituting Eq. (\ref{Psigp}) into Eq. (\ref{Ham}) and expanding $u_{\uparrow,\,
\downarrow}(x)$ and $v_{\uparrow,\, \downarrow}(x)$ in the
Bloch form in terms of $u_{l}$ and $v_{l}$  ($l$ labels the Bloch
eigenstate), we obtain the
Bogliubov-de Gennes (BdG) equations $M\Delta\phi=\omega\Delta\phi$, with
$\Delta\phi=\left(  u_{\uparrow l},v_{\uparrow l},u_{\downarrow l}%
,v_{\downarrow l}\right)  $ and  $\int
dx\left(  \left\vert u_{\uparrow l}\right\vert ^{2}-\left\vert v_{\uparrow
l}\right\vert ^{2}+\left\vert u_{\downarrow l}\right\vert ^{2}-\left\vert
v_{\downarrow l}\right\vert ^{2}\right)  =1$, and the matrix $M$ reads
\begin{widetext}
\begin{equation}
M=\left(
\begin{array}
[c]{cccc}%
L^{\left(  \uparrow\downarrow\right)  }\left(  k+q\right)  & c\phi_{\uparrow0}^{2} &
\Omega+c\phi_{\uparrow0}\phi_{\downarrow0}^{\ast} & c\phi_{\uparrow0}%
\phi_{\downarrow0}\\
-c\left(  \phi_{\uparrow0}^{\ast}\right)  ^{2} & -L^{\left(  \uparrow\downarrow\right)  \ast
}\left(  k-q\right)  & -c\phi_{\uparrow0}^{\ast}\phi_{\downarrow0}^{\ast} &
-\Omega-c\phi_{\uparrow0}^{\ast}\phi_{\downarrow0}\\
\Omega+c\phi_{\uparrow0}^{\ast}\phi_{\downarrow0} & c\phi_{\downarrow0}%
\phi_{\uparrow0} & L^{\left(  \downarrow\uparrow\right)  \ast}\left(  -k-q\right)  &
c\phi_{\downarrow0}^{2}\\
-c\phi_{\uparrow0}^{\ast}\phi_{\downarrow0}^{\ast} & -\Omega-c\phi
_{\downarrow0}^{\ast}\phi_{\uparrow0} & -c\left(  \phi_{\downarrow0}^{\ast
}\right)  ^{2} & -L^{\left(  \downarrow\uparrow\right)  }\left(  -k+q\right),
\end{array}
\right)  \label{BDG}%
\end{equation}
with
\begin{equation}
L^{\left(  m_{1}m_{2}\right)  }\left(  k\right)  =-\frac{1}{2}\left(  2in+ik\right)
^{2}+V_{0}\sin^{2}\left(  x\right)  -i\gamma\left(  2in+ik\right)
-\mu+2c\left\vert \phi_{m_{1}0}\right\vert ^{2}+c\left\vert \phi_{m_{2}0}\right\vert
^{2},
\end{equation}
\end{widetext}
and $m_{1},m_{2}=\uparrow,\downarrow$.  By solving the BdG equations numerically, the Bogoliubov excitation spectrum can be extracted. Note that, for a BEC trapped in optical lattices, two different types of instabilities of the BEC, {\it i.e.} dynamical instability and Landau instability \cite{InstabilityT},  can break the superfluidity of the model system, both of which have been extensively studied in theory \cite{InstabilityT} and experiments \cite{InstabilityE}. In this work, in order to avoid dynamical instability, which is relevant to our detailed calculations, we have limited ourselves to the stable parameter regime \cite{Zhang2013}. Then the dynamic structure factor can be found via \cite{LeiChenJLTP,Menotti} $S(q,\omega)=\sum_{j}Z_{j}(q)\delta(\omega-\omega_{j}(q))$,
where $Z_{j}(q)$ and $\omega_{j}(q)$ are the excitation strength and
frequency from the ground state to the $j-$th Bloch band, respectively.  In particular, the static structure factor for the model system can be immediately read off as \cite{LeiChenJLTP,Menotti}
\begin{equation}
S(q)=\sum_{j}Z_{j}(q). \label{Spo}%
\end{equation}

We present in Figs.  (\ref{band}) (a3)-(d3) the low-energy excitation spectrum of an optically trapped BEC with SOC corresponding to the four bands in Figs. \ref{band} (a1)-(d1), respectively. We find that, when the model BEC has an ordinary band, the familiar linear relation $\epsilon(q)\sim q$ arises (Fig. \ref{band} (a3)); whereas, remarkably, when the model BEC has a perfect flat band, a quadratic dispersion $\epsilon(q)\sim q^2$ emerges (Fig.  \ref{band} (d3)). Such distinct change in the excitation behaviour of the model system when the band flatness varies is also clearly observed in the dynamic structure, which is shown in Figs \ref{band} (a4)- (d4). In particular, the static structure factor (see black solid lines in Figs. \ref{band} (a4)-(d4)) exhibits a crossover from a linear relation $S({\bf q})\sim q$ to a quadratic relation $S({\bf q})\sim q^2$, when the band structure transforms from the ordinary into the perfect flat.

The crossover from the linear dispersion $\epsilon(q)\sim q$ to the quadratic dispersion $\epsilon(q)\sim q^2$ in the excitation spectrum of the model BEC \cite{Keeling2011,SuperExplain}, when the band structure transits from an ordinary band to a  locally perfect flat, can be intuitively understood in connection with the effective mass $m^*$ near the band minima.  As previously mentioned, a perfect flat band is associated with an almost infinite effective mass $m^*\rightarrow \infty$, therefore, the $q^2$ term is expected to varnish in the single-particle dispersion relation $\epsilon_0(q)$ (corresponding to zero kinetic energy) and the leading term can only emerge as $\epsilon_0(q)\sim q^{4}$. Thus, by using the above results $S(q)\sim q^2$ for a perfect flat band and Feyman's relation $\epsilon(q)=\epsilon_0(q)/S(q)$, we immediately have $\epsilon({ q})\sim q^2$ which explains the numerical results in Figs. \ref{band} (c1) and (d1). In contrast, in the opposite case of an ordinary band when $m^*\sim 1$, the single-particle kinetic energy is finite such that $\epsilon_0(q)\sim q^2$. Hence from $S(q)\sim q$ and  Feynman's relation, we have the familiar linear relation $\epsilon(q)\sim q$ in the BEC. We point out that, while the existence of flat bands in an optically trapped quasi-1D BEC with SOC can be described with a single-particle picture, the emerging quadratic low-energy dispersion when the band is perfectly flat is a many-body effect, which results from the interplay between the interaction and the band's flatness.

Finally, we have also analyzed how the excitation strength $Z_j$ is affected by the lattice strength $V_0$, in cases when the band is ordinary (Fig. \ref{band}(a1)) and when the band is flat (Figs. \ref{band}(b1)-(d1)), respectively.  As is clearly shown in Figs \ref{band} (a2)-(d2), where the first three Bogoliubov
bands are plotted, the dynamic structure factor is significantly affected by the optical lattice compared to the free-space case \cite{GPE3}. In particular, for a given value of momentum transfer $p$, it is possible to excite several
states corresponding to different bands. For example, Fig. \ref{band} (d2)  shows that when
 a density perturbation with $q=0.8k_L$ is generated in the BEC, not only
the first excitation strength $Z_1=0.37$ obtained, but also the second excitation $Z_2=0.04$.
An important consequence is that, on the one hand, it is possible to excite the high-energy states with small values of $p$; on the other hand, one can also excite the low energy states in the
lowest band with high momenta $p$ outside the first Brillouin zone. Such excitation behaviour is shared by both the ordinary band  (Fig. \ref{band} (b3)) and the flat bands (Figs. \ref{band} (b4)-(d4)) and therefore the existence of flat
bands cannot be revealed in the excitation strength $Z_j$ alone.

\section{Conclusion} \label{CO}

Overall, the crossover from linear to quadratic dispersion in the low-energy excitation spectrum presents a striking manifestation of the transition of an ordinary band into a perfect flat band, which permits the direct probe of flat band using the Bragg Spectroscopy. The experimental realization of our scenario amounts to controlling four parameters whose interplay underlies the physics of this work: the lattice strength $V_0$, SOC parameters $\gamma$ and $\Omega$, and the interatomic interaction $c$. All these parameters are highly controllable in the state-of-the-art technologies: $V_0$ can be changed from $0E_R$ to $32E_R$; both $\gamma$ and $\Omega$ can be changed by varying the angle between the two Raman lasers or through a fast modulation of the laser intensities \cite{Zhangyp2013}; in typical experiments to date, we can calculate the interaction coupling $c=0.05$ with the relevant parameters \cite{SOCBose} of $a= 100 a_{B} $ with $a_B$ the Bohr radius. Thus, we expect the phenomena discussed in this work be observable within the current experimental capabilities.

To conclude, we have found that the excitation spectrum of an optically trapped quasi-1D BEC with SOC alters significantly when the band flatness varies. In particular,  when the model BEC exhibits a perfect flat band (corresponding to $m^*\rightarrow \infty$ at the band minima),  a quadratic dispersion $\epsilon(q)\sim q^2$ emerges in the low-energy excitation spectrum; whereas, if the band is ordinary,  the familiar linear dispersion $\epsilon(q)\sim q$ arises. The variation in the flatness of band also alters the dynamic structure significantly.In particular, the static structure factor for a perfect band is quadratic in momenta $S(q)\sim q^2$, in contrast to the case of an ordinary band when $S(q)\sim q$ is linear. Based on these results, we propose to use Bragg spectroscopy to probe the arising flat band in an optically trapped quasi-1D BEC. The experimental verification of the new
dynamic features predicted in this work is expected to provide a significant
advance in our understanding of systems exhibiting flat-band-related phenomena.

\section{Acknowledgments}
We thank Biao Wu, Qizhong Zhu, Li Mao and P. Chatle for stimulating discussions. This work is supported by the
NSF of China (Grants No. 11274315 and No. 51331006). Y. Hu acknowledge support from Institut f\"ur Quanteninformation GmbH. Z.C. is supported by the NSFC(Grant No.11404026).

\cleardoublepage

\end{document}